\documentclass{ckm}                 % twocolumn proceedings style
\usepackage{txfonts}            % if you have it, to get Times family
\newcommand{\be}{\begin{equation}}
\newcommand{\ee}{\end{equation}}
\newcommand{\bea}{\begin{eqnarray}}
\newcommand{\eea}{\end{eqnarray}}
\confname{Workshop on the CKM Unitarity Triangle, IPPP Durham, April
  2003}
\title{Non factorizable effects in nonleptonic $B$ decays to charmonium}
\author{F. De Fazio\addressmark{a}}
\address[a]{Istituto Nazionale di Fisica Nucleare, Sezione di Bari, Italy}
\begin{document}
\begin{abstract}
We  discuss the validity of factorization for exclusive two-body B
to charmonium transitions. In particular, we consider the role of
non factorizable corrections in selected two-body modes.
\end{abstract}
%% \maketitle needs to be after the author and address info and the
%% abstract...
\maketitle
%% standard LaTeX from here on...
\section{Introduction}

Nonleptonic two-body $B$ decays play a crucial role in the study
of CP violation and in the determination of fundamental Standard
Model parameters, such as the Cabibbo-Kobayashi-Maskawa (CKM)
matrix elements or the angles of the unitarity triangle. However,
it is very difficult to deal with strong interaction effects for
these purely hadronic processes. One of the oldest prescriptions
to compute nonleptonic transitions is the factorization ansatz, of
which various formulations have been proposed. More recent
developments  exploit the presence of a large parameter, i.e. the
$b$ quark mass $m_b$. Since $m_b$ is much larger than the QCD
scale $\Lambda_{QCD}$, it is possible to evaluate the relevant
hadronic matrix elements as an expansion in the strong coupling
constant $\alpha_s(m_b)$ and in the ratio $\Lambda_{QCD}/m_b$.
Several approaches are based on such an expansion, such as QCD
factorization \cite{Beneke:1999br} or  perturbative QCD
\cite{Keum:2000ph}. Another possibility  is  to combine the heavy
quark expansion with non perturbative approaches such as QCD sum
rules \cite{Colangelo:2000dp} or to exploit the large energy
release to the final state in selected exclusive $B$ decays
\cite{Bauer:2000ew}. The realm of applicability of each method
should be assessed through the comparison with experimental data
in each specific case. In the following we consider $B$ to
charmonium transitions and compare available experimental data  to
theoretical predictions based on factorization. In this case, such
a comparison shows the existence of sizable violations. After a
brief review of the factorization approach, we analyse specific
decay modes for which the factorized amplitudes vanish;
nevertheless, they have been observed experimentally.

\section{B to charmonium transitions: factorization versus experimental data}

Let us consider a generic nonleptonic two-body B decay: $B \to M_1
M_2$. The effective hamiltonian describing this process is
obtained  integrating out the W field and can be written as:
$H_{eff}=\sum_i c_i(\mu) O_i$, where $c_i(\mu)$ are short distance
coefficients and $O_i$ local operators. As a consequence, the
relative amplitude reads as: \be A(B \to M_1 M_2)={G_F \over
\sqrt{2}} \sum_i \lambda_i c_i(\mu) \langle M_1 M_2|O_i(\mu) |B
\rangle \label{amplitude} \ee \noindent where $\lambda_i$
represents the product of the two elements of the
Cabibbo-Kobayashi-Maskawa matrix (CKM) involved in the considered
transition. Long distance effects are encoded in the matrix
elements of the operators $O_i$ and hence represent the non
perturbative ingredient of the calculation. In order to determine
such matrix elements, various approaches have been proposed. We
shall not review all of them here; instead, we  only consider the
factorization approach (in its various formulations) with
particular reference to final products consisting of a kaon and a
charmonium state.

The effective weak hamiltonian governing the process $B^- \to K^-
+ {\bar c}c$ is: \bea H_W&=&{G_F \over \sqrt{2}} \Big\{ V_{cb}
V_{cs}^* \left[ c_1(\mu) O_1 + c_2(\mu) O_2 \right] \nonumber
\\ &&- V_{tb } V_{ts}^* \sum_{i=3}^{10} c_i(\mu) O_i \Big\}
\label{heff} \eea \noindent where $O_1=({\bar c} b)_{V-A}({\bar
s}c)_{V-A}$ and $O_2=({\bar s} b)_{V-A}({\bar c}c)_{V-A}$ are
current-current operators, and $O_3-O_{10}$ are QCD and
electroweak penguin operators \cite{Buchalla:1995vs}. Let us
consider the mode $B^- \to K^- \chi_{c0}$, where $\chi_{c0}$ is
the $0^{++}$ state of charmonium. Naive factorization \cite{naive}
amounts to factorize the currents appearing in the $O_i$ and
computing the relevant amplitude inserting the vacuum in all
possible ways. As a consequence, one has: \bea A_{fact}(B^- \to
K^- \chi_{c0})={G_F \over \sqrt{2}} V_{cb} V_{cs}^* {\tilde c} \,
\langle K^-| ({\bar s}b)_{V-A}|B^- \rangle \times  \nonumber \\
\times \langle \chi_{c0} |({\bar c}c)_{V-A} |0 \rangle
\label{afact} \,,\eea \noindent where ${\tilde c}$ is a
combination of Wilson coefficients. Since the $\chi_{c0}$ is a
scalar ${\bar c}c$ particle, the latter matrix element in
(\ref{afact}) vanishes and therefore $A_{fact}=0$. Nevertheless,
both  Belle and BaBar Collaborations have reported observation of
this decay mode, giving:

\be {\cal B}(B^- \to K^- \chi_{c0})=(6.0^{+2.1}_{-1.8}\pm 1.1)
\times 10^{-4} \;\;\; \cite{Abe:2002mw}, \label{belledatum} \ee

\be {\cal B}(B^- \to K^- \chi_{c0})=(2.4\pm 0.7) \times 10^{-4}
\;\;\; \cite{Aubert:2002jn}. \label{babardatum} \ee

Such results clearly indicate that factorization is indeed
violated in the considered mode. Furthermore, Belle Collaboration
also provides the  ratio: \be {{\cal B}(B^- \to K^- \chi_{c0})
\over {\cal B}(B^- \to K^- J/\psi)} =(0.60^{+0.21}_{-0.18}\pm
0.05\pm0.08) \;\;\;, \label{belleratio} \ee \noindent showing that
the mode $B^- \to K^- \chi_{c0}$ proceeds with a rate comparable
to that of $B^- \to K^- J/\psi$.

A possible improvement of the naive factorization approach is
represented by the so called {\it generalized} factorization
\cite{generalized}, in which  the Wilson coefficients in the
factorized amplitude are treated as free parameters. If data are
available for some process, they can be fitted  and used as input
into other similar modes. However, it still holds that $A(B^- \to
K^- \chi_{c0})=0$ in generalized factorization.

It is interesting to consider a decay mode allowed in this
approximation, such as $B^- \to K^- J/\psi$. The factorized
amplitude reads: \bea A_{fact}(B^- \to K^- J/\psi)&=& 2{G_F \over
\sqrt{2}}V_{cb}V_{cs}^*a_2f_\psi M_\psi \times
\nonumber \\
&&\times F_1^{BK}(M_\psi^2) (\epsilon^* \cdot q) \label{psifact}
\,, \eea \noindent where $a_2=c_2+c_1/N_c$, $f_\psi$  is the
$J/\psi$ decay constant, $q$ the kaon momentum and $F_1^{BK}$ one
of the form factors parameterizing the matrix element $\langle K^-
|{\bar s} \gamma_\mu b| B^- \rangle$.

Using the form factor $F_1^{BK}$ computed in
\cite{Colangelo:1995jv} and ${\cal B}(B^- \to K^- J/\psi)=(1.01
\pm 0.05)\, 10^{-3}$ \cite{Hagiwara:fs}, one obtains:
$a_2^{eff}=0.38 \pm 0.05$. Scanning several form factor models,
the result would vary in the range: $a_2^{eff}=0.2 - 0.4$. The
obtained {\it effective} value of $a_2$ should be compared to the
QCD calculation, which gives $a_2^{NLO}(\mu=m_b)=0.163 \,
(0.126)$, in the naive dimensional regularization ('tHooft and
Veltman) scheme \cite{Buchalla:1995vs}. This comparison shows that
non factorizable effects are  sizable also in the case of $B^- \to
K^- J/\psi$.

A QCD improved factorization approach has been proposed for $B$
decays, exploiting the large value of $m_b$ \cite{Beneke:1999br}.
The approach holds for  $B \to M_1 M_2$ non leptonic decays when
$M_2$ is light,  $M_1$ being the meson picking up the spectator
quark in the decay. In this case, it has been shown that non
factorizable corrections are dominated by hard gluon exchanges,
while soft effects are confined to the $(BM_1)$  system. Naive
factorization is recovered at the leading order in $\alpha_s$ and
$\Lambda_{QCD}/m_b$. The approach does not hold when $M_2$ (the
emitted meson) is heavy, since a large overlap is expected between
$M_2$ and the $(BM_1)$ system. An exception is represented by the
emission of a quarkonium state, since, in the heavy quark limit,
its transverse size becomes small.

An analysis performed for $B^- \to K^- \chi_{c0}$ shows that
infrared divergences in the final result do not allow to apply
this method to such a process \cite{Song:2002mh}. As for $B^- \to
K^- J/\psi$, although the cancellation of infrared divergences has
been proven at the leading order in the  $1/m_b$ expansion, the
experimental data are not reproduced \cite{Cheng:2000kt}. In the
following section, we discuss the possibility that rescattering
diagrams, taking contribution from intermediate charmed mesons,
could play a role in the considered modes.

\section{Role of rescattering processes}

\begin{figure} \hbox to\hsize{\hss
\includegraphics[width=0.8\hsize]{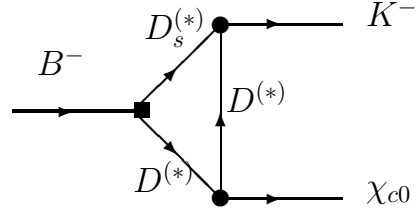}
\hss} \caption{Typical rescattering diagram contributing to $B^-
\to K^- \chi_{c0}$.} \label{triangolo}
\end{figure}

The decay $B^- \to K^- \chi_{c0}$ can be obtained by rescattering
of charmed intermediate states, as shown in fig. \ref{triangolo}
\cite{Colangelo:2002mj}. The decay is still induced by the
transition $b \to s {\bar c}c$ and the relevant CKM structure is
the same as for the direct transition.

A first analysis  of  rescattering diagrams of the kind shown in
fig. \ref{triangolo} is reported in \cite{Colangelo:2002mj} and
briefly summarized below.

The computation involves the weak matrix elements governing the
transitions $B \to D_s^{(*)}D^{(*)}$ and the strong couplings
between the charmed states $D_{(s)}^{(*)}D^{(*)}_{(s)}$ and the
kaon and the $\chi_{c0}$. There is experimental evidence that the
calculation of the amplitude by factorization reproduces the main
features of the $B \to D_s^{(*)}D^{(*)}$ decay modes
\cite{Luo:2001mc}. Therefore, neglecting the contribution of
penguin operators in (\ref{heff}), we can write: \bea \langle
D_s^{(*)-} D^{(*)0} | H_W | B^- \rangle &=& \displaystyle{G_F
\over \sqrt{2}}V_{cb}V_{cs}^* a_1 \langle D^{(*)0} | (V-A)^\mu |
B^- \rangle  \nonumber \\&& \langle D_s^{(*)-}| (V-A)_\mu | 0
\rangle \label{fact} \eea with $a_1=c_1+c_2/N_c$. In the heavy
quark effective theory, the matrix element $\langle D^{(*)0} |
(V-A)^\mu | B^- \rangle$ can be expressed in terms of a single
form factor, the Isgur-Wise function $\xi$ \cite{DeFazio:2000up},
while $\langle D_s^{*-}(p, \epsilon)| (V-A)_\mu | 0
\rangle=f_{D_s^*} M_{D_s^*} \epsilon^\mu $, where $f_{D_s^*}$ is
the $D^*_s$ leptonic constant and $\epsilon$ its polarization
vector; the analogous matrix element for $D_s$ involves the
corresponding constant $f_{D_s}$.

Other hadronic quantities involved in the calculations are the
strong couplings of a pair of charmed mesons to the kaon and to
the $\chi_{c0}$.

The $D_s^{(*)} D^{(*)} K$ couplings, in the soft $\vec p_K \to 0$
limit, can be related to a single effective constant $g$, as it
turns out considering the effective QCD Lagrangian describing the
strong interactions between the heavy mesons and the octet of the
light pseudoscalar mesons \cite{hqet_chir} $ {\cal L}_I = i \; g
\; Tr[H_b \gamma_\mu \gamma_5 {A}^\mu_{ba} {\bar H}_a]$, with $
{A}_{\mu ba}=\frac{1}{2}\left(\xi^\dagger\partial_\mu \xi-\xi
\partial_\mu \xi^\dagger\right)_{ba} $ and  $\xi=e^{i{\cal M} \over
f_\pi}$. The matrix ${\cal M}$ contains the fields of the octet of
the light pseudoscalar mesons. For example,  the $D^*_s D K$
coupling, defined through the matrix element \be <D^0(p)
K^-(q)|D^{*-}_s (p+q,\epsilon))>= g_{D^{*-}_s D^0 K^-} \, \,
(\epsilon \cdot q)   \label{gddk} \ee is related to the effective
coupling $g$ through the relation:
\begin{eqnarray}
g_{D_s^{*-} D^{0} K^-}&=& - 2 \sqrt{m_D m_{D_s^*}} {\displaystyle
g \over f_K}  \,\,\,\ . \label{gcoupl}
\end{eqnarray}

As for the coupling of the $\chi_{c0}$ state to a pair of
$D_{(s)}$ mesons, defined by the matrix element \be
 \langle
D^{0}_{(s)}(p_1) \bar D^{0}_{(s)}(p_2)| \chi_{c0} (p)\rangle=
g_{D_{(s)} D_{(s)} \chi_{c0}}\,, \label{gddchi} \ee \noindent an
estimate can be obtained considering the $D_{(s)}$ matrix element
of the scalar $\bar c c$ current: $\displaystyle \langle
D_{(s)}(\mathrm{v}^\prime) | \bar c c | D_{(s)}(\mathrm{v})
\rangle$, assuming the dominance of the nearest resonance, i.e.
the scalar $\bar c c$ state, in the
$(\mathrm{v}-\mathrm{v}^\prime)^2$-channel and using the
normalization of the Isgur-Wise form factor at the zero-recoil
point $\mathrm{v}=\mathrm{v}^\prime$. This allows us to express
$g_{D_{(s)} D_{(s)} \chi_{c0}}$ in terms of the constant
$f_{\chi_{c0}}$ that parameterizes the matrix element $\langle 0|
\bar c c | \chi_{c0}(q)\rangle= f_{\chi_{c0}} m_{\chi_{c0}}$. One
obtains:
\begin{equation}
g_{D_{(s)} D_{(s)} \chi_{c0}}= - 2  {\displaystyle m_{D_{(s)}}
m_{\chi_{c0}} \over f_{\chi_{c0}}} \,\,\,. \label{gchi}
\end{equation}
The method can also be applied to $g_{D^*_{(s)} D^*_{(s)}
\chi_{c0}}$. \par However, the determinations of the couplings
described above do not account for the off-shell effect of the
exchanged $D_{(s)}$ and $D^*_{(s)}$ particles,  the virtuality of
which can be large. As discussed in the literature, a method to
account for such effect relies on the introduction of form factors
$g_i(t)=g_{i0}\,F_i(t)$, with $g_{i0}$ the corresponding on-shell
couplings (\ref{gddk}), (\ref{gddchi}). A simple pole
representation for $F_i(t)$ is: $F_i(t)=\displaystyle{\Lambda_i^2
-m^2_{D^{(*)}} \over \Lambda_i^2 -t}$, consistent with QCD
counting rules \cite{Gortchakov:1995im}. The parameters in the
form factors represent a source of uncertainty in our analysis. In
the evaluation of the diagrams, we compute at first the absorptive
part, and then derive the real part through a dispersive
representation.

Before turning to the numerical analysis, it is worth considering
rescattering contributions of intermediate charm mesons to the
decay mode $B^- \to K^- J/\psi$. The hadronic information for
determining rescattering amplitudes are the same as for $B^- \to
K^- \chi_{c0}$, with the only difference in the strong
$D^{(*)}_{(s)} D^{(*)}_{(s)} J/\psi$ couplings that can be
expressed in terms of the parameter $f_{J/\psi}$, using the same
vector meson dominance method applied to derive eq. (\ref{gchi}).

We have now to fix the values of the various hadronic parameters.
The Wilson coefficient $a_1$, common to all the amplitudes, can be
put to $a_1=1.0$ as obtained by the analysis of exclusive $B \to
D^{(*)}_s D^{(*)}$ transitions. Moreover, we use $f_{D_s}=240$
MeV, in the range quoted by the Particle Data Group
\cite{Hagiwara:fs}, and $f_{D^*_s}=f_{D_s}$ consistently with our
approach  that exploits the large $m_Q$ limit. For the Isgur-Wise
universal form factor $\xi$, the expression $\displaystyle
\xi(y)=\Big({ 2 \over y+1}\Big)^2$ is compatible with the current
results from the semileptonic $B \to D^{(*)}$ decays.

A discussion is needed about the $D_s^{(*)} D^{(*)} K$ vertices.
For the effective coupling $g$  one can use the CLEO result
$g=0.59 \pm 0.01 \pm 0.07$ obtained by the measurement of  the
$D^*$ width \cite{Anastassov:2001cw}. Several estimates of $g$
have appeared in the literature; in particular, potential models
give values close to one \cite{Yan:gz}, while other determinations
point towards lower values \cite{g}. The discrepancy may been
attributed to relativistic effects \cite{Colangelo:1994jc}. The
value obtained by CLEO is in the upper side of the theoretical
calculations \cite{Colangelo:2000dp,becirevic}. We choose to be
conservative, and vary this parameter in the range: $0.35 < g <
0.65$ that encompasses the largest part of the predictions.

In (\ref{gchi}) we use  $f_{\chi_{c0}}=510\pm40$ MeV obtained by a
standard two-point QCD sum rule analysis \cite{Colangelo:2002mj}.
As for the couplings $D^{(*)}_{(s)} D^{(*)}_{(s)} J/\psi$,
expressions analogous to (\ref{gchi}) involve $f_{J/\psi}$, for
which we use the experimental measurement. Assuming that the
amplitude relative to $B^- \to K^- J/\psi$ deviates from the
factorized result because of the contribution of the rescattering
term: $\tilde {\cal A}_{exp} = \tilde {\cal A}_{fact} + \tilde
{\cal A}_{resc}$, one can constrain the values of $\Lambda_i$ for
the calculation of ${\cal B}(B^- \to K^- \chi_{c0})$. The result
is ${\cal B}(B^- \to K^- \chi_{c0})=(1.1-3.5) \times 10^{-4}$, to
be compared to (\ref{belledatum},\ref{babardatum}). \par \noindent
Two conclusions can be drawn from the present study: i)
rescattering amplitudes are  sizeble in $B^- \to K^- \chi_{c0}$
and in $B^- \to K^- J/\psi$; ii) they might explain the large
branching ratio observed for $B^- \to K^- \chi_{c0}$.

The calculation could be improved in several points: considering
additional intermediate states\footnote{The contribution of
orbital and radial excitations of the intermediate charmed mesons
is expected to be suppressed by smaller values of the leptonic
constants and of the strong couplings.}, using improved values for
the input parameters or inserting new experimental data. A more
refined analysis is indeed in progress.

It is also  interesting to estimate rescattering effects in other
channels which are also forbidden in the factorization
approximation, such as $B^- \to K^- \chi_{c2}$ or $B^- \to K^-
h_c$, $\chi_{c2}$ and  $h_c$ being the $2^{++}$ and  the $1^{+-}$
charmonium states, respectively. In particular, the latter decay
mode would also be interesting {\it per se}, since the $h_c$ has
been observed in $p {\bar p}$ annihilation, but it is not an
established particle yet \cite{Hagiwara:fs}. A preliminary result
gives ${\cal B}(B^- \to K^- h_c) \simeq (1-3) \times 10^{-4}$
\cite{noi} to be compared to the prediction for the inclusive mode
${\cal B}(B^- \to X h_c) \simeq (0.13-0.14)\%$
\cite{Beneke:1998ks}.

It has been suggested that a possible decay chain to observe this
mode could be: $B^- \to K^- h_c \,, h_c \to \gamma \eta_c $ with
$\eta_c \to K {\bar K} \pi$ or $\eta_c \to \eta \pi \pi$, and
${\cal B}(h_c \to \eta_c \gamma)\simeq 0.50 \pm 0.11$
\cite{Suzuki:2002sq}. Using such a prediction, our result would
give: ${\cal B}(B^- \to K^- h_c \to K \eta_c \gamma \to K (K {\bar
K}\pi) \gamma) \simeq(2.5 - 7.5 ) \times 10^{-6}$, suggesting that
this mode could be accessible at the currently operating
experimental facilities.

\section{Conclusions}
Present experimental results on two-body $B$ to charmonium
transitions show large non factorizable contributions. We
suggested that such contributions can be interpreted by
rescattering of intermediate charmed resonances. The numerical
evaluation of such effects  for $B^- \to K^- \chi_{c0}$ shows some
agreement with experimental data. Contributions of similar sizes
are expected for $B^- \to K^- h_c$. Predictions for other decay
modes could confirm this picture.

Partial support from the EC Contract No. HPRN-CT-2002-00311
(EURIDICE) is acknowledged. P. Colangelo and T.N. Pham are
thanked for fruitful collaboration on the topics discussed above.

\end{document}